\documentclass[12pt,twoside,letterpaper]{article}

\newcommand{\wlnu}{\ensuremath{\mbox{W}^{\pm} \rightarrow \ell^{\pm} \nu}}

\newcommand{\zll}{\ensuremath{\mbox{Z}^{\circ} \rightarrow \ell^+\ell^-}}
\newcommand{\zee}{\ensuremath{\mbox{Z}^{\circ} \rightarrow \mbox{e}^+\mbox{e}^-}}

\newcommand{\ppbar}{\ensuremath{p\overline{p}}}

\newcommand{\gamw}{\ensuremath{\Gamma \left( \mbox{W} \right) }}
\newcommand{\gamz}{\ensuremath{\Gamma \left( \mbox{Z} \right) }}

\newcommand{\gamwlnu}{\ensuremath{\Gamma \left( \wlnu \right) }}

\newcommand{\gamzll}{\ensuremath{\Gamma \left( \zll \right) }}
\newcommand{\sigw}{\ensuremath{\sigma_{\mbox{W}}}}
\newcommand{\sigz}{\ensuremath{\sigma_{\mbox{Z}}}}

\newcommand{\dzero}{\ensuremath{\mbox{D\O}}}

\renewcommand{\appendix}{
  \renewcommand{\section}{
      \newpage\thispagestyle{plain}
      \secdef\Appendix\sAppendix}
  \setcounter{section}{0}
  \renewcommand{\thesection}{\Alph{section}}
}

\newcommand{\Appendix}[2][?]{
  \refstepcounter{section}
  \addcontentsline{toc}{appendix}
      {\protect\numberline{\appendixname~\thesection} #1}
  {\flushleft\large\bfseries\appendixname\ \thesection\par

   \nohyphens\centering#2\par}
  \sectionmark{#1}\vspace{\baselineskip}
}

\newcommand{\sAppendix}[1]{
  {\flushleft\large\bfseries\appendixname\par
   \nohyphens\centering#1\par}
  \vspace{\baselineskip}
}

\newcommand{\nohyphens}{\hyphenpenalty=10000\exhyphenpenalty=10000\relax}

\newcommand{\etal} {{\em et al.}}

\usepackage{epsfig}

\usepackage{color}

\usepackage{ifpdf}
\ifpdf
 \usepackage{url,boxedminipage}
 \usepackage{graphicx,thumbpdf}
 \definecolor{rltred}{rgb}{0.75,0,0}
 \definecolor{rltgreen}{rgb}{0,0.5,0}
 \definecolor{rltblue}{rgb}{0,0,0.75}
 \usepackage[colorlinks%
  ,bookmarks%
  ,urlcolor=rltblue%
  ,filecolor=rltgreen%
  ,linkcolor=rltred%
  ,pdftitle={pdftitLe}%
  ,pdfsubject={pdfsubject}%
  ,pdfkeywords={pdfkeywords}%
  ,pdfauthor={Joe User <joeuser@fnal.gov>}%
  ,pdfpagemode={UseOutlines}%
  ,bookmarksopen=true%
  ,bookmarksnumbered=true%
  ,pdfstartview={Fit}%
  ]{hyperref}
 \DeclareGraphicsExtensions{.pdf,.jpg,.jpeg}            
\else
 \usepackage{graphicx}                                  
 \DeclareGraphicsExtensions{.eps,.ps,.eps.gz,.ps.gz}    
 \newcommand{\href}[2]{#2}                   

\fi 

\def\brzee{3.3658}
\def\brzeee{0.0023}

\newcommand{\enu}   {\mbox{${e\overline{\nu}_{e}}$}}
\newcommand{\mnu}   {\mbox{${\mu\overline{\nu}_{\mu}}$}}
\newcommand{\tnu}   {\mbox{${\tau\overline{\nu}_{\tau}}$}}
\newcommand{\Wtolnu} {\mbox{$\mathrm{W}\rightarrow\ell\overline{\nu}_{\ell}$}}
\newcommand{\Wtoenu}{\mbox{$\mathrm{W}\rightarrow\mathrm{e\overline{\nu}_{e}}$}}

\newcommand{\Wtoqqb} {\mbox{$\mathrm{W}\rightarrow q_{i}\overline{q}_{j}$}}
\newcommand {\GF}      {G_{\mathrm{F}}}
\newcommand {\MZ}      {M_{\mathrm{Z}}}
\newcommand {\MW}      {M_{\mathrm{W}}}

\newcommand {\GW}      {\Gamma_{\mathrm{W}}}
\newcommand {\qqb}        {q\bar{q}}

\newcommand {\udb}        {u\bar{d}}
\newcommand {\csb}        {c\bar{s}}

\newcommand {\alphamz}    {\alpha(\MZ)}
\newcommand {\alphasmz}    {\alpha_{\rm{s}}{\rm(\MZ)}}
\newcommand {\alphasmw}    {\alpha_{\rm{s}}{\rm(\MW)}}

\begin{document}

\thispagestyle{empty}

\begin{center}
  \begin{Large}
  {\bf Updated SM calculations of $\sigma_{W}/\sigma_{Z}$ at the Tevatron and 
the W boson width}
  \end{Large}
\end{center}

\vspace{0.5in}

\begin{center}
\href{mailto:p.renton1@physics.ox.ac.uk}{Pete Renton}
\footnote{p.renton1@physics.ox.ac.uk}\\ \vspace{1.0ex} 
\emph{University of Oxford, UK}
\end{center}

\begin{abstract}
The central value and theoretical uncertainties on the cross section 
ratio  $\sigma_{W}/\sigma_{Z}$ at the Tevatron are evaluated using
the NNLO calculations and the latest MSTW PDFs.

The partial width, total width and branching ratios of the W boson 
in the Standard Model, in the light of the 
latest electroweak calculations, are also updated.
\end{abstract}


\section{Introduction}

In this note we consider the theoretical uncertainty on the ratio of 
leptonic rates for the inclusive production of $W$ and $Z$ bosons at the
Tevatron ($\sqrt{s}$ = 1.96 GeV), 
namely,
\begin{eqnarray*}
\mbox{R} & = & 
\frac{\sigma \times \mbox{B}\left( \ppbar \rightarrow \wlnu \right)}
     {\sigma \times \mbox{B}\left( \ppbar \rightarrow \zll \right)}\, .
\end{eqnarray*}
from which the $W$ leptonic branching ratio and an indirect determination
of the total $W$ width can be extracted.

Published data on R come from Run~II measurements from 
the CDF collaboration \cite{cdfresult2}, together with those from 
both the CDF \cite{cdfresult1} and $\dzero$ \cite{d0result1} 
collaborations from Run~I. A note is in preparation of a Tevatron
combination of these published values.

From the definition of $R$, 
\begin{eqnarray*}
\mbox{R} & = & \frac{\sigma \cdot \mbox{B}(\wlnu)}
{\sigma \cdot \mbox{B}(\zll)} \\
 & = & \frac{\sigw}{\sigz} \cdot
\frac{\gamz}{\gamzll} \cdot \frac{\gamwlnu}{\gamw}\, ,
\end{eqnarray*}
we can extract the branching ratio of $\wlnu$, $\gamwlnu / \gamw$, by
using a Standard Model calculation for $\sigw / \sigz$ and the LEP
measurement of the $\zee$ branching ratio, 
namely  $B(\zee) = (\brzee \pm \brzeee) \%$,
assuming lepton universality~\cite{PDG}.

In a  previous Tevatron combination~\cite{rcomb_prelim} of 
preliminary Run~II results on R, together with those from Run~I, 
the results of the calculation in Ref.~\cite{willis} were used to
assign a theoretical uncertainty on the ratio $\sigw / \sigz$.
In that study a program based on the QCD NNLO expression developed by 
Van Neerven, \etal~\cite{vn 1, vn 2} was used
and gave the ratio of cross sections as $\sigw / \sigz$ = $3.361 \pm 0.024$.
However, the calculation was tree-level as far as electroweak
vertices are concerned. Consequently, there was an uncertainty
in the definition of $\sin^2\theta_W$, which was accounted for by
an additional uncertainty of $\pm 0.048$.  The value for the
cross section ratio used was  $\sigw / \sigz = 3.361 \pm 0.054$.
Subsequent updates of Ref.~\cite{willis} in Ref.~\cite{willis1}
gave, $\sigw / \sigz = 3.370$ with an uncertainty 
for the electroweak component alone of $\pm 0.014$. 
The CTEQ6.1 and MRST2001E PDF sets were used in these studies.

\section{Details of the calculation}

Recently updated sets of PDFs from the MSTW Collaboration (formerly MRST)
have been made available~\cite{MRST2006}. These new NNLO PDFs are 
interfaced to the NNLO program~\cite{stirling}, used to calculate
$\sigw, \sigz$ and  $\sigw / \sigz$, which is again 
based on the results of  Van Neerven, \etal~\cite{vn 1, vn 2}. 
The results presented here use this program, modified as discussed below.

The couplings of the Z boson to fermion-pairs have been changed from 
the Born-level formulation to using the effective couplings derived
from fits to LEP and SLD Z boson data; namely 
using $\rho$ = 1.0050 $\pm$ 0.0010 
and sin$^{2}\theta_{eff}$ = 0.23153 $\pm$ 0.00016~\cite{PDG}.

The Cabibbo-Kobayashi-Maskawa~\cite{cabibbo,km}(CKM) 
matrix elements used are also modified from the default values used
in the program.
The values used are the unconstrained measured values from~\cite{PDG}.
These values, rather than the unitarity constrained values are used because
the value of R can be used to give constaints on unitarity of the CKM
matrix and also to extract V$_{cs}$, which is poorly known from direct
measurement. The CKM values and uncertainties used are 
given in Table~\ref{tab:CKM}.

\begin{table}
\centering
\caption{
Values and uncertainties of the CKM matrix elements used and the
resulting uncertainty on X. The values are those not constrained by
Unitarity.
\label{tab:CKM}
}
\vspace*{0.1in}
\begin{tabular}{|l|l|l|l|}
\hline
 CKM element & value & uncertainty & $\Delta X$ \cr
\hline\hline
V$_{ud}$    &   0.97377   & 0.00027 & 0.0019 \cr
V$_{us}$    &    0.2257   & 0.0021  & 0.0016 \cr
V$_{ub}$    &   0.0043    & 0.0003  & 0.0000 \cr
V$_{cd}$    &   0.230     & 0.011   & 0.0039 \cr
V$_{cs}$    &   0.957     & 0.095   & 0.0050 \cr
V$_{cb}$    &   0.0416    & 0.0006  & 0.0000 \cr
\hline
\end{tabular}
\end{table}

The central value obtained is X = $\sigw / \sigz = 3.363$. The 
uncertainties on this which have been investigated are from 
a) PDF variations, b) uncertainties in the Z boson electroweak parameters, 
and c) uncertainties in the W boson CKM elements.

For the PDF unceratinties the eigenvector method was used. The values
of X = $\sigw / \sigz$ were computed for the 15 pairs of eigenvectors.
This gives 15 pairs of values $\Delta X_{up}$ and $\Delta X_{down}$,
corresponding to the ``up'' and ``down'' components of each pair. The
positive uncertainty on X was taken to be $\Delta X_{up}$ if 
$\Delta X_{up} >$ 0 and $\Delta X_{down} <$ 0. The positive uncertainty
on X was taken to be $\Delta X_{down}$ if $\Delta X_{down} >$ 0 and 
$\Delta X_{up} <$ 0 (and {\it vice versa} for the negative uncertainty).
In the case where both the ``up'' and ``down'' variations
are positive (negative) the value $\sqrt{(\Delta X_{up}^{2} + 
\Delta X_{down}^2)/2}$ was taken to be the positive (negative) 
uncertainty and the other component was set to zero. The positive and
negative components are then separately added in quadrature, giving
$\Delta X_{+}$ = 0.013 and $\Delta X_{-}$ = 0.010. We take the uncertainty
on X from the PDFs to be $\pm$ 0.013.

Note that the uncertainties in $\alphasmz$ and $\alphasmw$ are not
explicitly taken into account in the eigenvector method. However,
these are expected to largely cancel in the ratio considered 
here~\cite{MRST2006}. The value of the electromagnetic coupling
constant at the $\MZ$ scale, $\alphamz$, is not directly used in
these computations. Instead the values of $\GF$ and the vector boson
masses are used, thus absorbing some of the higher-order electroweak effects.
The widths of the W and Z bosons are also not used directly in the
cross section ratio calculation. This is, it is a zero-width approximation.
Again finite width effects are expected to largely cancel in the ratio,
but this has not explicitly been verified.

The uncertainty in the Z boson electroweak parameters 
$\rho$ = 1.0050 $\pm$ 0.0010 and sin$^{2}\theta_{eff}$ = 0.23153 $\pm$ 0.00016
were obtained by changing the values of each parameter in turn 
by $\pm 1 \sigma$ and adding the uncertainties in quadrature.
The result is $\Delta$X = $\pm$ 0.003.

For the CKM uncertainties each of the CKM elements in 
Table~\ref{tab:CKM} was moved by $\pm 1 \sigma$ and adding the 
uncertainties in quadrature. The result is $\Delta$X = $\pm$ 0.050.
The largest uncertainty comes from V$_{cs}$, which is poorly known
from direct measurement. This makes the theory estimate of X significantly
larger than previous estimates.

Combining these uncertainties in quadrature gives 

\begin{displaymath}
  X  = \sigw / \sigz = 3.363 \pm 0.052.
\end{displaymath}

\section{Branching ratios and widths of W boson in SM}

The W-boson decays weakly into either 
a quark-antiquark pair or a lepton and its corresponding neutrino.
The partial leptonic decay width is given by~\cite{rosner94}
\begin{equation}
    \Gamma(\Wtoenu) = \frac{\GF \MW^{3} }{ 6\pi\sqrt{2}}  
( 1 + \delta^{SM}_\ell) = 226.6 \pm 0.2 \hspace*{0.1cm} MeV. \ \ 
\end{equation}
The values $\GF$ = (1.16637 $\pm$ 0.00001) x 10$^{-5}$ GeV$^{-2}$
\footnote{Including the new result from the MuLan Collaboration~\cite{MuLan} gives
( $\GF$ = (1.166371 $\pm$ 0.000006) x 10$^{-5}$  GeV$^{-2}$. 
The FAST Collaboration~\cite{FAST} also have a new result, namely
$\GF$ = (1.166353 $\pm$ 0.000009) x 10$^{-5}$  GeV$^{-2}$. 
Using the updated world average
value from ~\cite{MuLan} gives neglible changes to the results reported here.}
and $\MW$ = 80.398 $\pm$ 0.025 GeV are used in the calculation.
The uncertainty is dominated by that in $\MW$. Note that 
by using the values of $\GF$ and $\MW$ to determine the SM value
of $\Gamma(\Wtoenu)$, the electroweak
corrections $\delta^{SM}_\ell$ are small ($\delta^{SM}_\ell$ = -0.34$\%$),
because the bulk of the corrections are absorbed in $\GF$ and $\MW$.

The partial width to $\qqb$ final states, for massless quarks, is given by
\begin{equation}
    \Gamma(\Wtoqqb) = f_{EW} f_{QCD} \Gamma(\Wtoenu) \mid V_{ij} \mid^{2}.  
\ \ 
\end{equation}
where f$_{EW}$ = (1 + $\delta^{SM}_q$) and $\delta^{SM}_q$ 
is the electroweak correction, 
with $\delta^{SM}_q$ = -0.40$\%$~\cite{rosner94}, 
and f$_{QCD}$ = 3(1 + $\alphasmw/\pi$ + 1.409($\alphasmw/\pi$)$^{2}$ + ...)
 is a QCD colour correction factor and V$_{ij}$  is the
CKM matrix element for i=u,d and j=d,s,b. 

The total width $\GW$ in the SM is given approximately by
\begin{equation}\label{eqn-gwsm}
    \GW = ( 3 + 2 f_{QCD} ) \Gamma(\Wtoenu) = 2.0932 \pm 0.0022 
\hspace*{0.1cm} GeV,  \ \
\end{equation}
where the uncertainty from $\alphasmw$ =0.1196 $\pm$ 0.0021
is 1.0 MeV, and that from $\MW$ is 2.0 MeV. The form in this
equation is approximate and neglects the differences in the
electroweak radiative corrections for leptons and quarks. This 
small effect is however included in the numerical value given.

From the above values the W leptonic branching ratio  
is computed to be
\begin{equation}\label{eqn-brsm}
 B(\Wtolnu) = (10.83 \pm 0.01)\%.
\hspace*{0.1cm} GeV.  \ \
\end{equation}

The CKM matrix elements entering into W decay are given in Table~\ref{tab:CKM}.
The main $\qqb$ decay modes are $\udb$ and $\csb$. The $\qqb$ branching ratio
thus gives mainly constraints on the matrix elements V$_{ud}$ 
and V$_{cs}$. Since the former is well known from other measurements, the
$\qqb$ mode can be used to give V$_{cs}$. Also the W leptonic 
branching ratio can be used to test the CKM unitarity constraint.




\section{Summary}

The Standard Model value of X, the ratio of the total W boson to Z boson cross
sections, has been estimated using the latest MSTW PDFs. An improved
electroweak formalism for the Z boson has been used and, for the W boson
production the latest direct CKM measurements have been used. 
The result is 
\begin{displaymath}
  X  = \sigw / \sigz = 3.363 \pm 0.052.
\end{displaymath}

Various properties of the W boson in the Standard Model have also been 
updated using revised electroweak corrections~\cite{rosner94}. 
The partial leptonic decay width is 
\begin{equation}
    \Gamma(\Wtoenu) = 226.6 \pm 0.2 \hspace*{0.1cm} MeV. \ \ 
\end{equation}

The total width $\GW$ is
\begin{equation}
    \GW = 2.0932 \pm 0.0022 
\hspace*{0.1cm} GeV,  \ \
\end{equation}
and the W leptonic branching ratio is computed to be
\begin{equation}
 B(\Wtolnu) = (10.83 \pm 0.01)\%.
\hspace*{0.1cm} GeV.  \ \
\end{equation}

\section{Acknowledgements}
I would like to thank W.J. Stirling and R.S. Thorne for providing their
latest PDFs and their NNLO program and for discussion on the cross section
ratio.

I would also like to thank J. Rosner and T. Takeuchi for updating their
Standard Model W boson properties and for valuable discussion.


\end{document}